# Contrasting to spark creativity in software development teams

**Marian Petre, The Open University, Milton Keynes, UK**
**Mary Shaw, Carnegie Mellon University, Pittsburgh, PA, USA**

**Abstract:**
Three decades of empirical research in high-performing software development teams provides evidence that creativity can be promoted by an effective, disciplined development culture. This paper describes *contrasting* as a key driver for creativity; describes *creativity moves*, tactics used by high-performing teams to produce useful contrasts; and characterizes key development behaviours observed to support a *culture* of creativity. The empirical research was carried out in a broad range of software development organizations and application domains.

Actionable insights:

- Great outcomes arise from commitment to the client's problem and its solution, not from use of particular tools or processes.
- The creative design mindset embraces reflection, re-assessment, skepticism, prioritising, and pragmatism. Creative design relies on breaking free of the seemingly obvious solutions.
- Developers intentionally generate contrasts to spark creativity, deliberately choosing different perspectives, different levels of abstraction, multiple design alternatives.
- Creativity requires critical consideration—with attention to gaps and contrary evidence.
- Documenting the net gains of creative behaviours in business terms can help influence management decisions.

## 1. INTRODUCTION

In recent discussions with numerous professional developers at industry conferences and in their workplaces, a recurrent theme has been the lack of investment in "the space for creativity". Developers have repeatedly pointed out that creativity is not just about radical innovations and new products. It's not just about "rockstar developers". Creativity is needed throughout software development, and throughout the software lifecycle.

There are many definitions of creativity. For this paper, 'creativity' combines some level of **novelty** with functionality that is **fit-for-purpose**. This resonates with Amabile's definition of creativity: *"the generation of a product that is judged to be novel and also to be appropriate, useful, or valuable by a suitably knowledgeable social group"* (Amabile 1982, p.1001). We interpret 'product' very broadly, as any element of software, from creating a new product line down to a single tweak. Similarly, we interpret 'novelty' broadly, as the addition of something new, different, refreshed at any level of the 'product'. We interpret 'fitness-for-purpose' in terms of suitability of the 'product' for serving its intended use for the intended users.

Creativity occurs in engineering as well as design, at many levels in software development: at scales from the design of innovative products to debugging and responses to pull requests; and beyond the software itself to the development environment (e.g., tools, practices). Vincenti (1993) distinguished between 'normal' design (applying previous designs to familiar problems) and 'radical' design (greenfield development, new concepts/products requiring major innovations). Simplistic interpretation of that distinction can imply (falsely) that creativity applies only to radical design. But creativity plays a role *throughout* development, e.g., when reviewing design documents, programming components, addressing bugs. In particular, creativity plays a role in challenging the limitations of normative problem solving, that is, 'reverting to what we usually do' (cf. Petre & Damian, 2014).

This paper describes:
- the evidence from professional software development practice that underpins this reflection (Section 2);
- *contrasting* as a key driver for sparking creativity (Section 3);
- *creativity moves*, tactics used by high-performing teams[1] to produce useful contrasts (Section 4); and
- the *culture* that supports creativity throughout software development (Section 5).

## 2. EVIDENCE

To address the developers' question "What can we do about promoting creativity?", this paper reflects on more than three decades of empirical research by one of the authors[2] in a broad range of more than 100 software development organisations and diverse application domains[3]. This research has taken many approaches, including:
- observing design meetings over weeks/months;
- shadowing experts (different projects, organisations, teams) over time as they work on problems and discuss them with colleagues;
- observing pair debugging in multiple companies/teams;
- collecting and analysing thousands of examples of design ephemera (e.g., back-of-an-envelope sketches, whiteboards, notes) from multiple teams/companies;
- conducting comparative studies with structured prompts;
- conducting quasi-experiments.

This paper reflects (across all of this research) on the development culture observed in high-performing software development teams recognised for their creativity. In particular, it characterises observed practices that promote creativity. Note that, from here, 'developers' refers to members of high-performing teams.

---

[1] 'High-performing teams here means teams that deliver working software on time, under budget, and without disastrous flaws.

[2] A partial bibliography is available at: https://softwaredesigndecoded.wordpress.com/annotated-bibliography/ - (Petre & van der Hoek, 2016)

[3] Application domains studied include: financial services, search engines/browsers, social networking, digital audio, digital video, computer games, automotive, aerospace, CAD systems, enterprise software, control systems, telecommunications, web development, software tools, government, medical information, retail, automation.

## 3. CONTRASTING SPARKS CREATIVITY

A key driver of creativity in high-performing teams is **contrasting**:  juxtaposing representations, ideas, perspectives, structures, design alternatives (etc.) and then playing them against each other through discussion and questioning.  Contrasting provides the basis for insight, through critical examination of the similarities and differences between the designs (or design options, or perspectives) being compared (cf., Polya, 1945). We use the verb, because contrasting is an active pursuit; it's not just about generating alternatives[4], but thinking critically about the contrasts between them and what they reveal. Contrasting provides the basis for thinking beyond the status quo, for challenging the current view of the problem-at-hand, and for escaping normative problem solving. Contrasting helps developers question the current thinking and design, and hence (by exposing assumptions, contradictions, omissions, possibilities, inefficiencies …) identify opportunities for creative improvement.

A key element of creativity in design expertise is attention beyond the design-at-hand. Developers challenge assumptions, models, designs.  Whereas many people look for evidence that things are working as expected, experts are more open to *contrary* evidence. They ask why; they engage users; they de-correlate – eliciting and contrasting different perspectives and hence often sparking creative insight.  They seek falsification, asking not just 'How would I know if this is right', but 'How would I know if this were wrong?' – and 'How would I know if an alternative were right?'  In doing so, they are mitigating against biases in human reasoning. For example, in both naturalistic and lab studies of test case selection (Teasley, Leventhal & Rohlman, 1994), testers were four times more likely to choose positive test cases than negative ones – four times more likely to try to show that the program works, than that it fails. Yet points of failure often reveal new design opportunities.

Developers tend to focus on the crucial, difficult, or unknown elements of the problem (which are often where the creativity is needed/arises), rather than leaping to familiar solutions based on superficial similarity.  This doesn't mean that developers don't use familiar solutions – but they do so after they have invested in critical analysis of the problem and context at hand, and they may do so creatively, with novel adaptations that address the particular context and hence improve fitness-for-purpose.  *Fitness-for-purpose* provides the evaluative focus that informs contrasting and design decision-making.  It is evident in the engagement with the domain and with stakeholders (including end-users) throughout the design process.  Curtis, Krasner, and Iscoe (1988), in reporting an empirical study of a large development team, quote a system engineer: "Writing the code isn't the problem; understanding the problem is the problem".  Their study identifies domain knowledge as one of, if not the primary, factor related to the success of a software system. The contrast between the domain and software perspectives is rich ground for creativity.

---

[4] Here 'alternatives' refers not just to design alternatives, but perspectives, representations, levels of abstraction, contexts/domains, outputs of different tools – things whose critical comparison yields information that might inform the design, provide insight, or spark a creative leap.

A key observation is that expert use of contrasting is *deliberate*. Developers understand which perspective they're applying; they juxtapose contrasting perspectives and move deliberately between them to reveal and evaluate different aspects and assumptions. They do this continually throughout software design and development, not just in early design, and not just at the whiteboard.

High-performing teams contrast tools as well as solutions, often playing tools (e.g. notations, representations, processes, analysis tools…) against each other to increase gap detection. One developer explained: "Often, it's the mishmash of different ways of thinking that gets you the answer." A single representation will never be able to capture everything there is to know about a software system. Developers use notations as lenses, using whichever notation best suits the task-at-hand, and swapping between notations to change perspective – and then contrasting between the notations to spark creativity.

Multiple techniques/tools imply more distinct ways to think. However, they also entail greater cognitive overheads, and they require intelligent coordination. So, the selection is not arbitrary; teams try tools, assess their merits, and assemble 'toolkits' that both fit their development culture and promote contrasts. They choose the tools that support both their creativity and productivity.

## 4. CREATIVITY MOVES: tactics for producing [useful] contrasts

Developers have a variety of tactics for seeking useful contrasts and challenging their thinking –thereby getting more traction to find effective solutions. We term these tactics/mechanisms **'creativity moves'**: deliberate identification/selection of things to contrast usefully and hence drive critical and creative design thinking. This section describes four categories of tactics: resonance, dissonance, expanding the space, limiting the space (Figure 1).

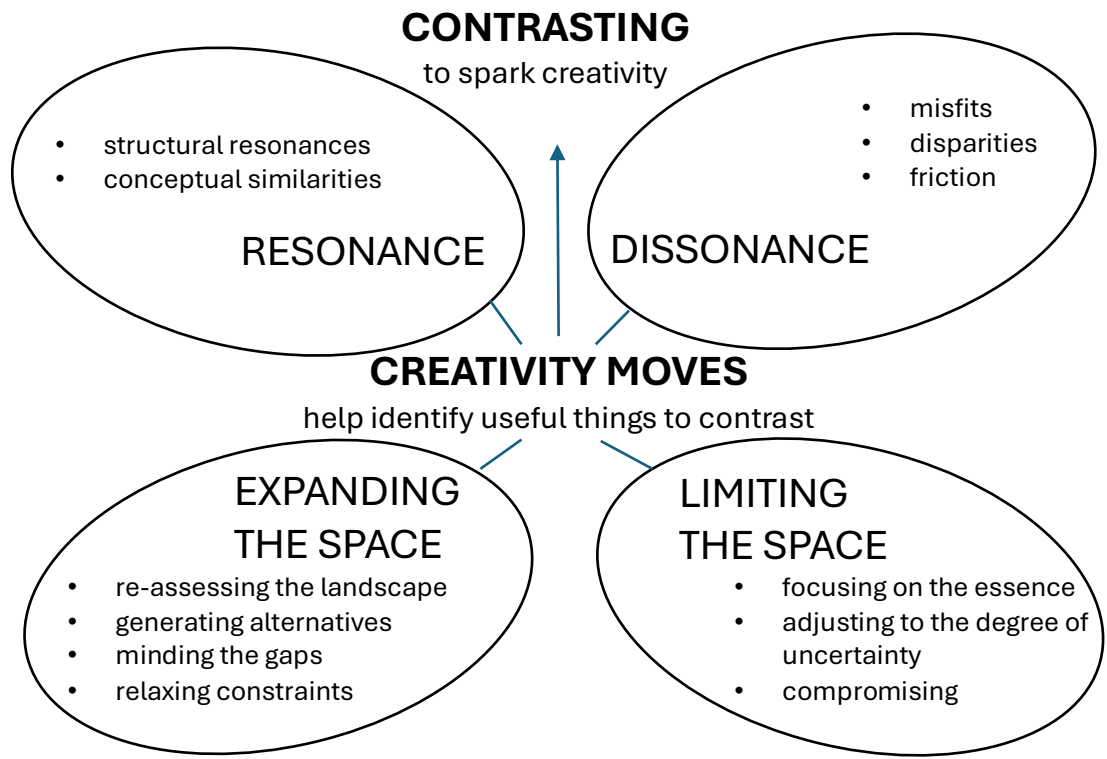

Figure 1: Overview of creativity moves

## 4.1 Resonance

One creativity move is **seeking structural resonances** across projects, across domains, sometimes through metaphors and analogies. As per Polya (1945, pp.10-11): Do you know a related problem? One with similar givens? Similar unknowns? Can you use the solution? The derivation? More broadly, if our design has structural or conceptual similarities to another design, what does our knowledge of that design suggest that we might re-apply? What properties, dependencies, constraints, assumptions might we look for?

One example of resonance emerged from observation of a software developer known for generating algorithms[5]. He was wrestling with a scheduling problem for parallel processing in digital audio. He walked around the corridors musing, until he noticed the exhaust manifold from a racing car propped in the hallway. Having stared at it, he headed quickly to his office to start coding, because he had solved his problem. The insight was that the pipes in an exhaust manifold take different routes around the car, but are all the same length, and all reach the same destination. He used that structural resonance to solve his scheduling problem in a way that provided a novel approach to the essence of the problem.

## 4.2 Dissonance

Another creativity move is **seeking dissonance**: attending to misfits and asking questions such as: What isn't consistent or compatible within the design, or between the problem and the proposed solution (i.e., dissonance-with-purpose)? What's missing? Is contrary evidence being brushed aside? Is the fit with purpose not quite right? Are exceptions rationalized away rather than addressed?

---

[5] The examples come from professional designers observed in industry unless a specific academic citation is given.

One developer recounted: "[An] example is to come up with designs that, on purpose, clash with existing patterns in the architecture.  It is then possible to evaluate if it clashes because it is generally not a good design, or … because the current architecture is lacking."  e.g., "… we identified that, in a REST system we need not just the location of an entity or its unique identifier but both. The location to get more info, if needed, the identifier because one entity might be in multiple locations (due to migrations).  Current services only stored location, not identifier.  The designs were often incompatible, but…"

Developers seek and/or apply dissonance in order to establish the solution-to-problem fit.  Seeking dissonance aims to uncover, use, and find things that are not aligned/congruent – and sometimes identify things that are missing.  Dissonance-with-purpose often arises from fixation on a known solution (or on doctrine) which then leads to inattention to the differences in the current problem; hence critical attention to dissonance can avoid issues and potentially lead to creative insight that goes beyond the status quo.

### 4.3 Expanding the space

Another creativity move is **expanding the space** of possibilities under consideration: Am I solving the right problem?  Is there another way to model or implement this? What if I generalized the problem? Do components really share the same assumptions? Are too many special cases being ignored? Would reframing the problem from a different viewpoint lead to a better solution?  One developer reflected: "I like to ask the question: What if we had to extend on X … or build feature Y on this design?"

**Re-assessing the landscape** is a way of examining barriers, understanding constraints, revealing assumptions, looking beyond the immediate issues, or broadening the problem definition in a way that provides insight and overcomes flaws.  Developers do this periodically throughout the design and development process, not just during early design[6].

**Generating alternatives** is one way to expand the space. One developer described his deliberate generation and comparison of alternatives to spark creativity and justify change: "It has become absolutely necessary for me to take the time to
(1) create a design space for architecture features, and
(2) compare architectures in this space based on their software quality attribute measures. By creating different dimensions in the space (such as types of database, security types, and message protocols) and then sampling this space by actually creating the various architectures, I have been able to get leadership to question where they are going architecturally and change it …"

**Minding the gaps** is another example of expanding the space.  Developers pay attention to the spaces between things, e.g., to interfaces, interactions between components, integration with other systems, domain concepts hidden behind standard data types, etc.

---

[6] This is at odds with many software development methodologies, which focus on convergence to a solution. So, sometimes experts step away from a methodology.

They pay attention to what isn't shown, to what's missing – whether from the design, or the information, or the tool they're using. This is often characterised as 'white space finding': examining a whole process (perhaps one that is not yet achievable) and trying to identify inefficiencies, obstacles, or gaps in the currently available products or services. This is one use of scenarios and use cases. One developer explained: "What I've found useful is use-case driven development: What demo shall we make that will be great? What's the gap between what we have and what we want? … It's like a forcing function … pressing us make our system to scale, to be more efficient, to be more beautiful in terms of the interface. … It's a way of stressing … and growing a system."

For tough problems, developers may use **counterintuitive acts** such as trying the opposite, or relaxing constraints (cf. deBono, 1992). This may not lead to a satisfactory solution, but it challenges their understanding and almost certainly leads to insights that help them free their thinking and identify creative possibilities.

### 4.4 Limiting the space

Complementary to expanding the space is **limiting the space**: "If you can't solve the problem, solve part of the problem." "Keep only one part of the condition, drop the other part." (Polya, 1945, p. 23) What could I do without? This resonates with the notions of "creativity of constraints" or "psychology of limitations": there is evidence that reducing the degrees of freedom, or introducing constraints temporarily, can spark creative thinking, because with fewer obstacles and fewer parameters, mental capacity is freed for insight and experimentation. Limiting the space is sometimes applied to the problem, and sometimes used to simplify the solution.

One developer described his design strategy of **focusing on the essence** and **deferring** non-critical issues: "We're developing an integration framework for composing AI and widgets for users. So the user can choose what kind of AI you want, what kind of widgets you want, update the dashboard live … I know that, in the long term, I need some sort of persistence mechanism. I could invest a lot in some some huge, complex database, or I could start out with a file because that's not the problem that I need to address yet. I need to address being able to say what sort of AI I want declaratively, and be able to spin it up on a machine somewhere. … But it's enough to start addressing this larger problem that needs to be decided before I talk about persistence mechanisms. … Some pieces need to be addressed first, but you need some scaffolding to even start. Once one piece solidifies, it informs the other pieces. … Then at the end hopefully you haven't compromised anything, and you have accomplished more than you might have hoped to."

Developers **adjust to the degree of uncertainty**, e.g., understanding the normal operating range. An example from scientific software: "When assessing the stability of a meteorological model, it doesn't matter if the temperature in Greenland is over $100°C$." Hence, developers are attentive to contextual factors that moderate uncertainty.

Similarly, developers have been seen to **compromise** in order to move work along. For example, for a reliability problem, recognizing that the cost of detection/repair is lower than the cost of prevention may make switching from prevention to reaction/remediation an effective compromise (Shaw, 2024). Developers deliberately make sub-optimal choices calculated to serve immediate needs (i.e., 'satisficing') – allowing them time to gather more information or evidence. In the meantime, they free themselves to brainstorm about the core and to try new approaches un-bogged-down by detail. In this way, satisficing frees space for invention and information gathering. In this way, creativity, as observed, also works with pragmatism.

## 5. THE CULTURE:  Mindset, dialogues, and socially-embedded practices

As discussed, developers use a repertoire of creativity moves that can generate contrasts and help to pivot the problem or the solution.  The creativity moves are part of a **design creativity mindset**, embracing:

- reflection – including analysis and review, taking time to look and think, engagement with problem *and* solution;
- re-assessment – consideration of the problem, the goal, and change over time;
- skepticism – critique, critical thinking, attention to contrary evidence;
- prioritising, and hence pragmatic deferring and compromising – doing *something* (even if inadequate) in order to make progress or gain insight;
- a principal commitment to fitness-for-purpose.

A key observation is that expert use of contrasting and creativity moves is *deliberate*.  It's easy for an individual to lapse; a team culture that embeds practices such as contrasting and creativity moves takes the onus off any one developer.  For example, when high-performing teams are in a meeting, there's often the 'guru in the corner', an experienced designer who will interrupt the discussion at crucial moments to ask the contrasting question about whether some proposal would work in a given use case, or if it is consistent with other software in the product line, or whether an alternative approach would be simpler or more efficient, etc.  Creativity is best supported by collaborative team behaviours.

### 5.1 Dialogues

Developers talk about the importance of 'Getting out of one's own head'.  High performing teams understand the value of dialogues, including dialogues with users and stakeholders, with other developers, and with themselves.  Dialogues provide a change of perspective, additional (sometimes unexpected) input, and a requirement to externalise thinking – all fuelling contrasts and contributing to creativity.

**Organised dialogues** include things like stand-ups, reviews, and design discussions.  They can provide coordination across the participants, input from different perspectives, and learning from experience.  As one developer reported about a personnel database system: "We had a problem where we lost data between sites.  So the team got together and, in reviewing the architecture, discovered that a couple of people had slightly different conceptions of the interactions between some of the core components.  In systematically discussing and resolving the misconceptions, we had a 'eureka moment' about some of the interactions that not only solved the problem but inspired a leaner, more efficient design."

**Consulting:** Developers don't expect to know everything – and know that they don't. They deliberately involve others both inside and outside their team (including stakeholders and users) when they have a purpose for doing so, often to obtain specialized technical or domain knowledge. As one developer said: "The realisation is: you don't know everything; there's probably someone who might." Richer information and contrasting perspectives contribute to creative insight.

**Spontaneous dialogues** often begin between two developers, then grow when others hear the conversation. For example, high-performing teams do 'pair de-bugging': developers sit together and talk through code, often deliberately with people who know different parts of the code base. This brings a fresh perspective to the code, and has a tendency to expose assumptions, misconceptions, and miscommunications. Similarly, many design problems are solved when developers bump into each other 'at the coffee machine', when unexpected input provides leverage on the problem-at-hand. Many companies recognise this and invest in shared facilities – and expect key individuals to spend time in those spaces so they are available to the dialogues that spark creativity.

**Externalising thought** (e.g., sketches and other representations, spoken descriptions) can support a form of dialogue with oneself or with the design (Schön's "conversation with the materials", 1983). As one developer explained: "... a developer talks through a problem with a colleague, or with an inanimate proxy [cf. 'rubber ducking']. Externalising the issue, explaining it to someone else, can make the reasoning more explicit and concrete, often breaking the unproductive cycle and providing insight." Externalising thought can help expose assumptions, omissions, and misconceptions. Similarly, representations on a whiteboard can expose thinking to external scrutiny, prompting dialogues with (and insights from) others (Cherubini et al., 2000; Petre et al., 2012, Petre, 2010).

## 5.2. The cultural essentials of creativity in SE

The interplay *between developers* plays a crucial part both in nurturing creativity and innovation (on the one hand) and in embedding systematic practice and rigour (on the other) – and hence in enhancing productivity overall. The team culture provides a safety net for detecting and addressing issues, by building trust and communication, and leveraging individual strengths and multiple perspectives,. The safety net gives developers license to be creative.

These are the basics of a creative SE culture, based on extensive observations of high-performing teams:
- **primary focus:** fitness-for-purpose; keeping an eye on the design goals;
- an alignment between business and engineering priorities, reflected in adequate **team autonomy** (as well as the commitment to fitness-for-purpose);
- **investment (**by management/organisation and team) in developers and creativity (e.g., in alternatives, learning, tools, dialogues, reflection);
- **valuing contribution**, rather than relying on metrics;
- **socially-embedded practices** that encourage questions and ideation, nurture trust and psychological safety, and embed systematic analysis and rigorous evaluation.

Creativity is empowered by socially-embedded and reinforced practices and values. As one developer expressed it: "You've got to actually care (not just go through the motions)". High-performing teams typically include someone who can translate between the engineering and business perspectives, conveying the business value of focusing on the goals (prioritising fitness-for-purpose) and valuing contribution (rather than just output or metrics). In demonstrating the business value, they justify investing in the space for creativity.

## 6. 'WHAT CAN WE DO?'

Developers often ask what they can do to promote creativity – particularly in an organization whose management is not aligned with this mindset.

The culture of creativity is not 'all or nothing'. Even high-performing teams that are largely autonomous run into periods of intense production and tight deadlines which can override the creative practices – temporarily. In such teams, the culture is renewed as soon as the deadlines and intensity ease. Sometimes it starts with pair debugging before release, when the value of critical dialogues is re-experienced. Often it starts with a retrospective on the product just delivered, which takes the team back into dialogues, contrasts, and reflection. Then other creative practices can be refreshed.

One way forward is for teams to make an internal commitment to a culture of dialogues and contrasts. With persistence, the behaviors can be embedded gradually over time. As the benefits of regular dialogues become evident, the team is more likely to renew its commitment to them, and to consider further investment. The team will build a repertoire of practices that fit them, and then will induct new members into that community of practice. For example, one organisation understood these benefits and used pair debugging and collaborative code review as onboarding mechanisms. The recruit was assigned a variety of pull requests that covered issues distributed around the code base. In addressing the requests, the recruit engaged in -debugging with different members of the team and became familiar with both the nature of the code base and the expertise in the team – all in the course of doing 'real work'.

It pays for developers to take initiative in making the alignment between business and engineering goals explicit – hence working toward a regular dialogue between the perspectives. For example, one team justified its investment in secure coding training and tools in terms of the significant reduction of the number of repeated vulnerabilities. The net gains of creative behaviour and culture can influence management over time – and potentially change company culture, making time and space for creativity. Documenting the benefits of being creative contributes credibility and justifies investment to repeat the behaviour.

## 7. CONCLUSION

For creativity to flourish, it must be embedded in a fertile culture. High-performing teams use systematic, disciplined practices that are socially embedded and reinforced. Importantly, because there is a disciplined culture, they are able to rely on the team to catch slips – thereby giving individuals the freedom to be creative and experiment without fear, and thereby justifying investment in dialogues, reflection, learning, tools – and creativity.

The team culture matters: it embodies the mindset that embraces multiple perspectives, engages with contrary evidence, and reinforces practices (such as contrasting, prioritization, design dialogues) that routinely challenge understanding and assumptions. This helps strengthen and develop the team *as well as* improving the software. It makes space for and fuels creativity. Without that culture, teams can stagnate. Without that culture, groups of high-performing *individuals* may not cohere into a high-performing *team*. Software expertise doesn't happen by accident, and neither does creativity in software design; these are practices that can be understood and nurtured. By making space in organizational culture, and by investing time for this mindset, these practices, and these dialogues, teams are making space for creativity to work and grow.

## ACKNOWLEDGEMENTS

We are grateful to all the industry developers who participated in studies and provided examples, and to helpful colleagues including Jakob Durstberger, Julian Harty, Jonathan Bailey, Mike Hoye, and Greg Wilson. We also thank the theme issue panel, including the anonymous referees, and Andre van der Hoek, who provided invaluable guidance.

## Authors

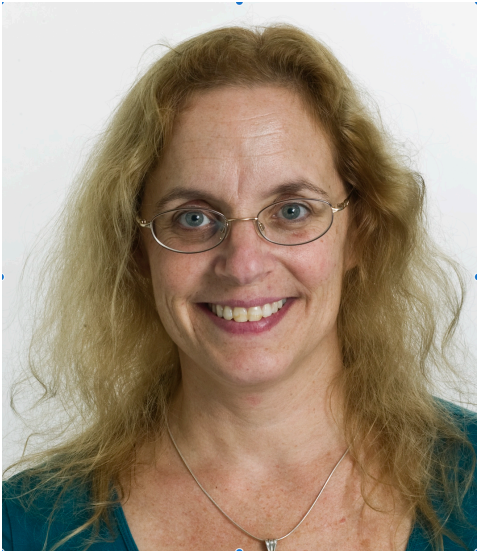

Marian Petre

Marian Petre is an Emeritus Professor at The Open University in the UK.  Her research on how software developers reason and communicate about design and problem solving is grounded in empirical studies of professionals in industry and draws on cognitive and social theory.  She received a Royal Society Wolfson Research Merit Award in recognition of her work and is a Fellow of the BCS and a Distinguished Scientist of the ACM.  Contact her at m.petre@open.ac.uk.

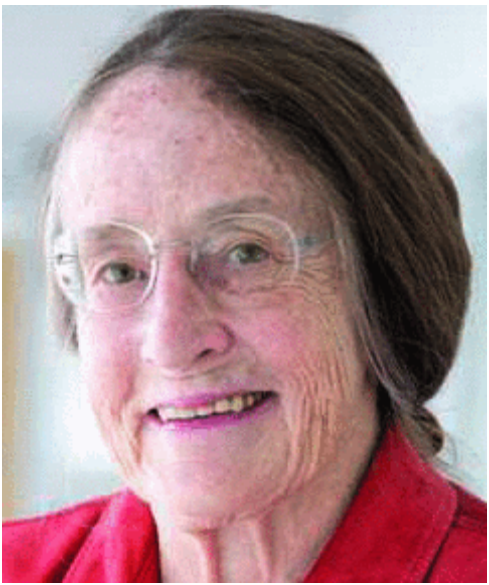

<u>Mary Shaw</u>

Carnegie Mellon University, Pittsburgh, USA

Mary Shaw is the Alan J. Perlis University Professor of Computer Science at Carnegie Mellon University, Pittsburgh, PA 15217 USA. Her research focuses on software engineering, particularly software architecture and the design of systems used by real people. She is a recipient of the US National Medal of Technology and Innovation and a Fellow of ACM, IEEE, and the AAAS. Contact her at mary.shaw@cs.cmu.edu.